\documentstyle[12pt,epsf]{article}

\newcommand{\be}{\begin{equation}}

\newcommand{\ee}{\end{equation}}

\begin{document}

\begin{titlepage}
\rightline{\vbox{\halign{&#\hfil\cr
&UQAM-PHE-98/08\cr
&UMN-TH-1730/98\cr
&TPI-MINN-98/26\cr
&December 1998\cr}}}
\vspace{0.5in}
\begin{center} 

\Large\bf   
The limits on CP-odd four-fermion operators containing strange quark field 
\\ 
\medskip 
\vskip0.5in 
 
\normalsize {{\bf C. Hamzaoui}$^a$\footnote{hamzaoui@mercure.phy.uqam.ca} and  
{\bf  
M. Pospelov}$^b$}\footnote{pospelov@tpi1.hep.umn.edu} 
 
\smallskip 
\medskip 
 
{\bf a}{ \sl  D\'{e}partement de Physique, Universit\'{e} du Qu\'{e}bec {\`a}  
Montr\'{e}al\\  
C.P. 8888, Succ. Centre-Ville, Montr\'{e}al, Qu\'{e}bec,  
Canada, H3C 3P8}  
 
\smallskip 
 
{\bf b}{ \sl Theoretical Physics Institute, 
 431 Tate Laboratory of Physics\\ 
Minneapolis, MN 55455 
USA} 
\end{center} 
 
\vskip1.0in

\noindent 
{\large\bf Abstract} 
\smallskip\newline 
The bounds on the neutron electric dipole moment and T-odd nucleon-nucleon  
interaction are used to extract the limits on the effective CP-odd  
four-fermion  
operators containing strange quark field. This completes the study of the  
dim=5,6 
CP-odd operators built from the light-quark fields. The limits are very strong 
  and comparable to 
those obtained previously for operators containing up and down  flavors.  
We also  
analyze the 
shift of the axionic vacuum, 
$\theta_{eff}$, induced by four-fermion operators in the presence of PQ  
mechanism and 
conclude that this gives subleading contributions to CP-odd observables as  
compared with the 
direct ones.  
\end{titlepage} 
 
\baselineskip=20pt 
 
\newpage \pagenumbering{arabic} 
 
\section{Introduction} 
 
The remarkable accord of standard model (SM) predictions with experiment 
does not remove the question of a more fundamental theory which would 
comprise SM as the low energy limit. Among various ways to look for the 
signs of a new theory, precise tests of fundamental discrete symmetries play 
an important role. 
 
The nil results for the electric dipole moments of the neutron, heavy atoms 
and diatomic molecules \cite{nEDM,mEDM,eEDM,molEDM} put in general very 
strong constraints on the CP-violating sector of a new theory and probe the 
energy scales unaccessible for direct observations at colliders \cite{KL}. 
Regardless what the particular construction for the new theory is, its 
relevant contribution at 1 GeV can be reexpressed in terms of effective 
operators of different dimensions suppressed by corresponding power of a 
high scale $M$ where these operators were generated:  
\begin{equation} 
{\cal L}_{eff}=\sum_{n\geq 4}\frac{c_{ni}}{M^{n-4}}{\cal O}_{i}^{(n)}, 
\end{equation} 
Here ${\cal O}_{i}^{(n)}$ are operators of dimension $n$ and $i$ stands for 
their different field content, Lorentz structures etc. Fields, relevant for 
low-energy dynamics, are gluons, three light quark fields, $u$, $d$ and $s$, 
and the electromagnetic field. The specifics of a given model enters only 
through the value of the coefficients $c_{ni}$ 
 
Dim=3, 4 operators can be combined to form $\theta $-term. In the absence of 
axion relaxation mechanism this operator is normally the most important on 
account of possible tree-level contributions $\sim $ 1. If PQ mechanism is 
operative $\theta \sim $ 1 is removed but $\theta $-parameter still can gain 
a nonzero value induced at low energy by CP-odd operators of bigger 
dimension. Dim=5 operators, which are usually suppressed by an additional $%
m_{q}/M$ ratio, are electric and chromoelectric dipole moments of quarks. 
Due to this additional mass ratio, these operators are suppressed by a large 
scale exactly as dim=6 operators built from four quark fields or purely from 
gluons (Weinberg operator). Most of the operators have been extensively 
studied in the literature \cite{KKY,KK,KL} and limited from experiment using 
PCAC and QCD sum rules techniques. For the operators with strange quarks, 
however, only the analysis of the chromoelectric dipole moment is available  
\cite{KKZ}. Recently some of the four-fermion operators with $s$-quark 
induced at a high scale in $SU(3)\times SU(2)\times U(1)$-symmetric form 
were limited using the fact these operators can be mixed with electric 
dipole moment operators for $u$ and $d$ fields at one-loop level \cite{HM}. 
 
In this letter we combine PCAC approach and the experimental bounds on 
neutron, mercury and thallium EDMs to put the limits on four-fermionic 
operators containing strange quark field. This completes the study of the 
relevant operators dim=6 and can be used for any model where these operators 
are generated. Another issue addressed here is the shift of axion vacuum by 
effective CP-odd four-fermion operators. This shift, $\theta_{eff}$ is 
estimated within the same approach . 
 
\section{CP-odd operators containing strange quark} 
 
In what follows, we adopt the classification of operators proposed in Refs.  
\cite{KKY,KK,KL}. Among the flavour-conserving CP-odd operators dim=6  
\begin{eqnarray} 
&&\kappa _{1}\frac{G}{\sqrt{2}}(\bar{s}s)(\bar{q}i\gamma _{5}q);\;\;\;\kappa 
_{2}\frac{G}{\sqrt{2}}(\bar{s}i\gamma _{5}s)(\bar{q}q);\;\;\;\kappa _{3}%
\frac{G}{\sqrt{2}}(\bar{s}i\gamma _{5}s)(\bar{s}s);  \nonumber \\ 
&&\kappa _{4}\frac{G}{\sqrt{2}}(\bar{s}t^{a}s)(\bar{q}i\gamma 
_{5}t^{a}q);\;\;\;\kappa _{5}\frac{G}{\sqrt{2}}(\bar{s}i\gamma _{5}t^{a}s)(%
\bar{q}t^{a}q);\;\;\;\kappa _{6}\frac{G}{\sqrt{2}}(\bar{s}i\gamma 
_{5}t^{a}s)(\bar{s}t^{a}s);  \nonumber \\ 
&&\kappa _{7}\frac{G}{\sqrt{2}}\frac{1}{2}\epsilon _{\mu \nu \alpha \beta }(%
\bar{s}\sigma _{\mu \nu }s)(\bar{q}\sigma _{\alpha \beta }q);\;\;\;\kappa 
_{8}\frac{G}{\sqrt{2}}\frac{1}{2}\epsilon _{\mu \nu \alpha \beta }(\bar{s}%
\sigma _{\mu \nu }t^{a}s)(\bar{q}\sigma _{\alpha \beta }t^{a}q); \\ 
&&\kappa _{9}\frac{G}{\sqrt{2}}(\bar{s}s)(\bar{e}i\gamma _{5}e);\;\;\;\kappa 
_{10}\frac{G}{\sqrt{2}}(\bar{s}i\gamma _{5}s)(\bar{e}e);\;\;\;\kappa _{11}%
\frac{G}{\sqrt{2}}\frac{1}{2}\epsilon _{\mu \nu \alpha \beta }(\bar{s}\sigma 
_{\mu \nu }s)(\bar{e}\sigma _{\alpha \beta }e).  \nonumber  \label{eq:oper} 
\end{eqnarray} 
we take those, containing $s$-quark field. We will numerate these operators $%
{\cal O}_{i}$ according to the constant $\kappa _{i}$ standing in front of 
them. $G$ in these formulae is Fermi constant. 
 
As an experimental input we use the following limits, obtained for the 
neutron EDM \cite{nEDM},  
\begin{equation} 
d_{N}<10^{-25}e\cdot cm, 
\end{equation} 
and mercury EDM \cite{mEDM} experiments:  
\begin{equation} 
d_{Hg}<9\cdot 10^{-25}e\cdot cm. 
\end{equation} 
The latter translates into the limit on the Schiff moment of $^{199}$Hg 
nucleus and eventually leads to the following bound on the effective 
CP-violating $\pi ^{0}$pp coupling \cite{mEDM,KL}:  
\begin{equation} 
\bar{g}_{\pi pp}<2\cdot 10^{-11}. 
\end{equation} 
 
To evaluate the contribution of operators (\ref{eq:oper}) to the effective 
coupling $\bar g_{\pi pp}$ we use the same method proposed earlier in Refs.  
\cite{KKY,KK,KL}. The operator ${\cal O}_1$ is the simplest in this respect. 
Using the PCAC reduction of the soft-pion field and calculating subsequent 
commutators, we reduce the contribution of the $(\bar s s)(\bar q i\gamma_5 
q)$ operator ${\cal O}_1$ to the matrix element of $\bar s s \bar q q $ 
operator over the proton:  
\begin{equation} 
\langle p\pi^0|\bar s s\bar d i\gamma_5 d|p\rangle=\frac{1}{f_\pi} \langle 
p|\bar s s\bar d d|p\rangle 
\end{equation} 
This matrix element can be estimated using vacuum insertion approximation:  
\begin{equation} 
\langle p|\bar s s \bar d d|p\rangle\simeq \langle 0| \bar q 
q|0\rangle\langle p|\bar s s +\bar q q|p\rangle\simeq5(1+\beta)\bar pp, 
\end{equation} 
where we take $\langle p| \bar d d|p\rangle\simeq\langle p| \bar u 
u|p\rangle= \langle p| \bar q q|p\rangle\simeq 5 \bar pp$. The analysis of 
the barion mass splitting and experimental data on pion-nucleon scattering 
suggest that the coefficient $\beta$, $\beta=\langle p| \bar 
ss|p\rangle/\langle p| \bar qq|p\rangle$, numerically is close to 0.6 \cite 
{KKZ,Z}. Thus, CP-violating coupling constant $\bar g_{\pi pp}$ is:  
\begin{equation} 
\bar g_{\pi pp}=5(1+\beta)\frac{\langle 0| \bar q q|0\rangle}{f_\pi} G\simeq 
8\cdot 10^{-6}\kappa_1.  \label{eq:direct} 
\end{equation} 
 
The rest of the four-quark operators give suppressed contributions to the 
effective T-violating nucleon-nucleon interaction. They either contribute 
only in $\eta$-exchange channel or do not work in vacuum factorization 
approach. To get the limits on these operators we use the neutron EDM bound. 
The neutron EDM can be induced as a result of the chiral loop, Fig. 1, where 
CP-violation resides in one of the meson-nucleon vertecies. In the limit of 
exact chiral symmetry this loop is logarithmically divergent in the infrared  
\cite{CDVW} which justifies its appearance in the chiral theory. For our 
purposes we choose $\Sigma^-K^+$ loop \cite{KKZ} where the operators 
containing $s$-quark will most likely contribute. In the real life, chiral 
symmetry is broken and the mass of kaons is rather large, so that estimated 
limit on $n\Sigma^-K^+$ coupling has rather large uncertainty:  
\begin{equation} 
g_{n\Sigma^-K^+}<2\cdot 10^{-11} 
\end{equation} 
Most of the quark operators from the set (\ref{eq:oper}) induce this 
coupling; to calculate their effect on it we use the same method, PCAC and 
vacuum factorization. Thus, for example, $\bar si\gamma_5s \bar dd$ operator 
contributes in the CP-odd vertex of interest in the following way:  
\begin{eqnarray} 
\langle \Sigma^-K^+|\bar si\gamma_5s\bar dd|n\rangle=\frac{i}{f_K}%
\langle\Sigma^-|\bar dd\bar s u|n\rangle \simeq \frac{i}{f_K}\langle 0| \bar 
q q|0\rangle\langle \Sigma^-|\bar s u|n\rangle \simeq  \nonumber \\ 
\frac{i\langle 0| \bar q q|0\rangle}{f_K}\frac{m_\Sigma-m_N}{m_s}\bar 
\Sigma^- n \simeq \frac{i\langle 0| \bar q q|0\rangle}{f_K}1.3 \bar \Sigma^- 
n  \label{eq:ksn} 
\end{eqnarray} 
and the $SU(3)$-octet type of splitting in the barion mass spectrum was used 
in the second line of Eq. (\ref{eq:ksn}). 
 
The limits on semileptonic operators, ${\cal O}_{9}$, ${\cal O}_{10}$ and $%
{\cal O}_{11}$, can be obtained from the limits on T-odd nucleon-electron 
interaction \cite{KL} by simply taking matrix elements from their 
strange-quark part over the nucleon. For the case of ${\cal O}_{9}$ and $%
{\cal O}_{10}$ these matrix elements can be easily obtained within the same 
PCAC approach. In the case of ${\cal O}_{11}$, however, the tensor charge of 
the strange quark over the nucleon, $\langle N|\bar{s}\sigma _{\mu \nu 
}s|N\rangle $, is not known. It is not reducible to the $s$-quark spin 
content over the nucleon, as it was asserted in Ref. \cite{EF}, because the 
latter is expressed by absolutely independent quantity $\langle N|\bar{s}%
\mbox{$\gamma_{\mu}$}\mbox{$\gamma_{5}$}s|N\rangle $. Moreover, unlike 
axial-vector operator, tensor operator is odd under charge conjugation and 
we expect the effects of strange and anti-strange quarks to cancel each 
other in the first approximation. The model calculations and lattice 
simulations for tensorial charges give indeed a very suppressed value for 
the strange quark contribution \cite{tcharge}. The same refers, of course, 
to the strange quark EDM operator, as it was discussed in \cite{HPR}. 
 
The resulting limits on the coefficients are summarized in Table 1. One can 
easily see that the best sensitivity is for ${\cal O}_{1}$ and ${\cal O}_{9}$ 
operators where $s$-quark enters only as $\bar{s}s$ and does not take part 
in spin dynamics. 
 
\section{Effective theta-term induced by CP-odd four-fermion operators} 
 
In all known models with significant amount of CP-violation in the 
flavor-conserving channel, the operators of dim$>$4 are usually accompanied 
by a large contribution to the theta term. In other words, $\theta _{loop}$ 
is usually by far more sensitive to the new CP-violating physics because it 
corresponds to the operator dim=4 and therefore need not experience scale 
suppression of the order $(\Lambda _{QCD}/M)^{2}$. Thus the CP-violating 
operators of dim$>$4 generated at a scale $M\sim M_{W}$ and higher are 
important only in the case when $\theta $-term is removed by an axion 
mechanism. We also assume here the existence of PQ mechanism. In the absence 
of CP violation, non-removable by PQ transformation, PQ symmetry sets theta 
parameter to zero \cite{PQ}. The situation is different in the presence of 
extra CP-violating sources, communicated by the operators dim$\ge 5$. These 
operators ${\cal O}_{i}$ shift the axion vacuum and generate additional 
indirect contribution to all CP-odd observables through the effective $%
\theta $-term given by the ratio of two correlators:  
\begin{eqnarray} 
\theta _{eff} &=&-\frac{K_{i}}{|K|},\;\;\mbox{where}\;\;K=i\left\{ \int 
dxe^{ikx}\langle 0|T(\frac{\mbox{$\alpha$}_{s}}{8\pi }G\mbox{$\tilde{G}$}(x),%
\frac{\mbox{$\alpha$}_{s}}{8\pi }G\mbox{$\tilde{G}$}(0))|0\rangle \right\} 
_{k=0} \\ 
K_{i} &=&i\left\{ \int dxe^{ikx}\langle 0|T(\frac{\mbox{$\alpha$}_{s}}{8\pi }%
G\mbox{$\tilde{G}$}(x),{\cal O}_{i}(0))|0\rangle \right\} _{k=0}.  \nonumber 
\end{eqnarray} 
Here $G_{\mu \nu }^{a}\mbox{$\tilde{G}$}_{\mu \nu }^{a}$ is abbreviated as $G%
\mbox{$\tilde{G}$}$. The calculation of $K$ is based on the use of the 
anomaly equation and the saturation of subsequent correlators by light 
hadronic states \cite{SVZ}:  
\begin{equation} 
K=\frac{m_{\pi }^{2}f_{\pi }^{2}m_{u}m_{d}}{2(m_{u}+m_{d})^{2}}. 
\end{equation} 
($F_{\pi }$ we use throughout the paper is 130 MeV). The same technique can 
be exploited in the case of $K_{1}$. For the case of chromoelectric dipole 
moments the explicit derivation of $K_{i}$ can be found in Ref. \cite 
{BU,Posp}. A similar type of calculation could be done for the most part of 
the four-fermion operators discussed here and in earlier works \cite{KKY, KK}%
. Using the anomaly equation in the form  
\begin{eqnarray} 
\partial _{\mu }\frac{m_{d}m_{s}\bar{u}\gamma _{\mu }\gamma _{5}u+m_{u}m_{s}%
\bar{d}\gamma _{\mu }\gamma _{5}d+m_{u}m_{d}\bar{s}\gamma _{\mu }\gamma _{5}s%
}{m_{s}m_{d}+m_{s}m_{u}+m_{d}m_{u}} =  \nonumber \\ 
\frac{2m_{u}m_{d}m_{s}}{m_{s}m_{d}+m_{s}m_{u}+m_{d}m_{u}}(\bar{u}\gamma 
_{5}u+\bar{d}\gamma _{5}d+\bar{s}\gamma _{5}s)+\frac{\mbox{$\alpha$}_{s}}{%
4\pi }G\mbox{$\tilde{G}$},  \label{eq:anom} 
\end{eqnarray} 
we apply standard technique of current algebra. The correlators of interest,  
$K_{i}$, can be rewritten in the form of the equal-time commutator, which 
for all sets of four-fermion operators we can calculate easily, plus the 
term containing the singlet combination of pseudoscalars built from quark 
fields. Thus, for ${\cal O}_{1}$ operator we have the following expression:  
\begin{eqnarray} 
K_{1} =\kappa _{1}\frac{G}{\sqrt{2}}\langle 0|\frac{m_{d}m_{s}(\bar{u}u)(%
\bar{s}s)}{m_{s}m_{d}+m_{s}m_{u}+m_{d}m_{u}}+\frac{m_{u}m_{d}(\bar{u}i%
\mbox{$\gamma_{5}$}u)(\bar{s}i\mbox{$\gamma_{5}$}s)}{%
m_{s}m_{d}+m_{s}m_{u}+m_{d}m_{u}}|0\rangle + \\ 
\int d^{4}x\langle 0|T\{\frac{im_{u}m_{d}m_{s}}{%
m_{s}m_{d}+m_{s}m_{u}+m_{d}m_{u}}(\bar{u}\gamma _{5}u+\bar{d}\gamma _{5}d+%
\bar{s}\gamma _{5}s)(x),{\cal O}_{1}(0)\}|0\rangle  
\nonumber  
\label{eq:k1} 
\end{eqnarray} 
The second line here is suppressed by an extra power of light quark masses 
in the numerator. It would bring a comparable contribution, though, if there 
were an intermediate hadronic state with a mass, vanishing in the chiral 
limit $m_{i}\rightarrow 0$. At the same time, the flavor structure of this 
term shows that the lightest intermediate state here is $\eta ^{\prime }$ 
which is believed to remain heavy even if quark masses vanish. Thus the 
contribution from the second term is negligible in the limit $m_{\pi }\ll 
m_{\eta ^{\prime }}$. The second term in the first line of Eq. (\ref{eq:k1}) 
is suppressed by $m_{u}/m_{s}$ ratio and effectively we get the following 
formula for the induced theta term due to the operator ${\cal O}_{1}$:  
\begin{equation} 
\theta _{eff}=-\kappa _{1}\delta _{1}\frac{G}{\sqrt{2}}m_{u}^{-1}
\langle 0|\bar{q}q|0\rangle ,  \label{eq:th} 
\label{eq:TEFF}
\end{equation} 
where $\delta _{1}$ is the ratio of the four-quark condensate to the square 
of $\bar{q}q$ condensate.  
\begin{equation} 
\delta _{1}=\frac{\langle 0|\bar{u}u\bar{s}s|0\rangle }{\langle 0|\bar{q}%
q|0\rangle ^{2}}. 
\end{equation} 
In the case of ${\cal O}_{1}$ we can use vacuum factorization and estimate 
that $\delta _{1}\sim 1$. For some of four-quark operators vacuum 
factorization does not work and we expect $\delta _{i}$ to be smaller than 1.
The appearence of $m_u^{-1}$ in Eq. (\ref{eq:TEFF}) is because the operator 
${\cal O}_1$ breaks chirality. Any answer for CP-odd observables, iduced by
$\theta$ will not contain this singularity. 
 
The value of $\theta _{eff}$ induced by ${\cal O}_{1}$ leads to an 
additional contribution to $\bar{g}_{\pi pp}$ constant,  
\begin{equation} 
\bar{g}_{\pi pp}(\theta _{eff})=\frac{m_{u}m_{d}}{m_{u}+m_{d}}\frac{\sqrt{2}%
\theta _{eff}}{f_{\pi }}=\delta _{1}\kappa _{1}\frac{G\langle 0|\bar{q}%
q|0\rangle }{f_{\pi }}\frac{2m_{d}}{m_{u}+m_{d}}=1.3\cdot 10^{-6}\kappa _{1}, 
\end{equation} 
which should be compared with the direct contribution (\ref{eq:direct}). We 
see that in the case of ${\cal O}_{1}$ the indirect contribution related to 
theta term gives 15-20\% correction to the CP-odd vertex $\bar{g}_{\pi pp}$. 
In fact, this is the biggest value of $\theta _{eff}$, generated by the set 
of operators (\ref{eq:oper}). This is especially true for the operators $%
{\cal O}_{3}$ and ${\cal O}_{6}$ composed exclusively from strange quark 
which induce $\theta _{eff}$ with an additional parametrical suppression $%
(m_{u}^{-1}+m_{d}^{-1})/m_{s}$. We have also done similar calculation for 
the operators composed from the $u$ and $d$ field \cite{KKY,KK,KL} and the 
result for the $\theta $-driven contributions never exceeds 20\% from the 
direct contribution. The estimates for $\theta _{eff}$ are included into 
Table 1. 
 
\section{Conclusions} 
 
We have considered the limits on the four-fermion CP-odd operators 
containing strange quarks coming from the neutron, thallium and mercury EDM 
experiments. This completes the study of CP-odd dim=5,6 operators built from 
light-quark fields. We observe that the limits are very strong, especially 
for the operators $\bar{s}s\bar{q}i\mbox{$\gamma_{5}$}q$ and $\bar{s}s\bar{e}%
i\mbox{$\gamma_{5}$}e$ where strange quark is in some sense ''spectator''. 
The limits summarized in Table 1 are almost as strong as for operators 
composed from $u$ and $d$ quarks. This is because the strange quark 
condensate is the same as for up and down quarks in the flavor $SU(3)$ 
symmetry approximation and the content of the $s$-quark in the nucleon in 
scalar, pseudoscalar and axial-vector channels is also significant. The 
limits on the operators ${\cal O}_{1},$ ${\cal O}_{9}$ and ${\cal O}_{10}$ 
are extracted with much smaller error than those for the rest of the 
operators. Other limits are estimated within an order of magnitude, mainly 
because of the chiral loop, used to obtain the EDM of the neutron. The 
infrared logarithm enhancement factor, $\log {m_{K}}$, is numerically 
important only in the limit $m_{K}\rightarrow 0$. 
 
There is an alternative method of limiting four-fermion operators used in 
Ref. \cite{HM}. In this work different linear combination of operators (\ref 
{eq:oper}) were taken to form a different set, invariant under standard 
model group. At one-loop level some of these operators can be mixed with 
EDMs of $u$ and $d$ quarks with the coefficients proportional to $m_{s}\log 
(\Lambda /1\mbox{GeV})$. The comparison of the limits obtained in Ref. \cite 
{HM} and in the present work shows that they are complementary. For most of 
the operators the limits obtained here are stronger, although ${\cal O_{7}}$ 
and especially ${\cal O_{11}}$ can be better constrained from their one-loop 
mixing with quark and electron EDM operators. 
 
The shift of the axionic vacuum induced by four-fermion operators is shown 
to give contributions to CP-odd observables normally smaller than the direct 
contributions not associated with $\theta _{eff}$. For some operators this 
type of correction can be as large as 15-20\% from the direct contribution. 
 
{\bf Acknowledgments} 
 
M.P. would like to thank P. Herczeg and I. B. Khriplovich for valuable 
discussions. This work is supported in part by N.S.E.R.C. of Canada and DOE 
grant DE-FG02-94ER-40823 at the University of Minnesota.

\newpage 
 
\begin{center} 
Table 1. 
\end{center} 
 
The limits obtained on the $CP$-odd four-fermion operators from neutron, 
thallium and mercury EDM experiments. 
 
\vspace{1cm} 
 
\begin{center} 
$ 
\begin{tabular}{|c|c|c|c|c|c|c|} 
\hline 
&  &  &  &  &  &  \\  
& ${\cal O}_{1}$ & ${\cal O}_{2}$ & ${\cal O}_{3}$ & ${\cal O}_{4}$ & ${\cal %
O}_{5}$ & ${\cal O}_{6}$ \\  
&  &  &  &  &  &  \\ \hline 
&  &  &  &  &  &  \\  
$\kappa _{i}$ & $3\cdot 10^{-6}$ & $7\cdot 10^{-5}$ & $7\cdot 10^{-5}$ & $%
3\cdot 10^{-4}$ & $3\cdot 10^{-4}$ & $3\cdot 10^{-4}$ \\  
&  &  &  &  &  &  \\ \hline 
&  &  &  &  &  &  \\  
$|\theta _{eff}|$ & $5\cdot 10^{-5}\kappa _{1}$ & $5\cdot 10^{-5}\delta 
_{2}\kappa _{2}$ & $2\cdot 10^{-6}\kappa _{3}$ & $5\cdot 10^{-5}\delta 
_{4}\kappa _{4}$ & $5\cdot 10^{-5}\delta _{5}\kappa _{5}$ & $5\cdot 
10^{-7}\kappa _{6}$ \\  
&  &  &  &  &  &  \\ \hline 
\end{tabular} 
$ 
 
\vspace{1cm} 
 
$ 
\begin{tabular}{|c|c|c|c|c|c|} 
\hline 
&  &  &  &  &  \\  
& ${\cal O}_7$ & ${\cal O}_8$ & ${\cal O}_9$ & ${\cal O}_{10}$ & ${\cal O}%
_{11}$ \\  
&  &  &  &  &  \\ \hline 
&  &  &  &  &  \\  
$\kappa_i$ & $3 \cdot 10^{-5}$ & $2 \cdot 10^{-5}$ & $2\cdot10^{-7}$ & $%
10^{-5}$ & $10^{-5}$ \\  
&  &  &  &  &  \\ \hline 
&  &  &  &  &  \\  
$|\theta _{eff}|$ & $5\cdot10^{-5}\delta_7\kappa_7$ & $5\cdot10^{-5} 
\delta_8\kappa_8$ & - & - & - \\  
&  &  &  &  &  \\ \hline 
\end{tabular} 
$ 
\end{center} 
 
\newpage 
 
{\bf {\large {Figure captions.}}} 
 
{\normalsize Chiral loop diagrams, inducing the EDM of the neutron. Dirac 
structure of $CP$-violating vertex is proportional to 1. This diagram 
logarithmically diverges in the limit $m_{K}\rightarrow 0$. } 
 
{\normalsize \vspace{3cm}  
\begin{figure}[hbtp] 
\begin{center} 
{\normalsize \mbox{\epsfxsize=144mm\epsffile{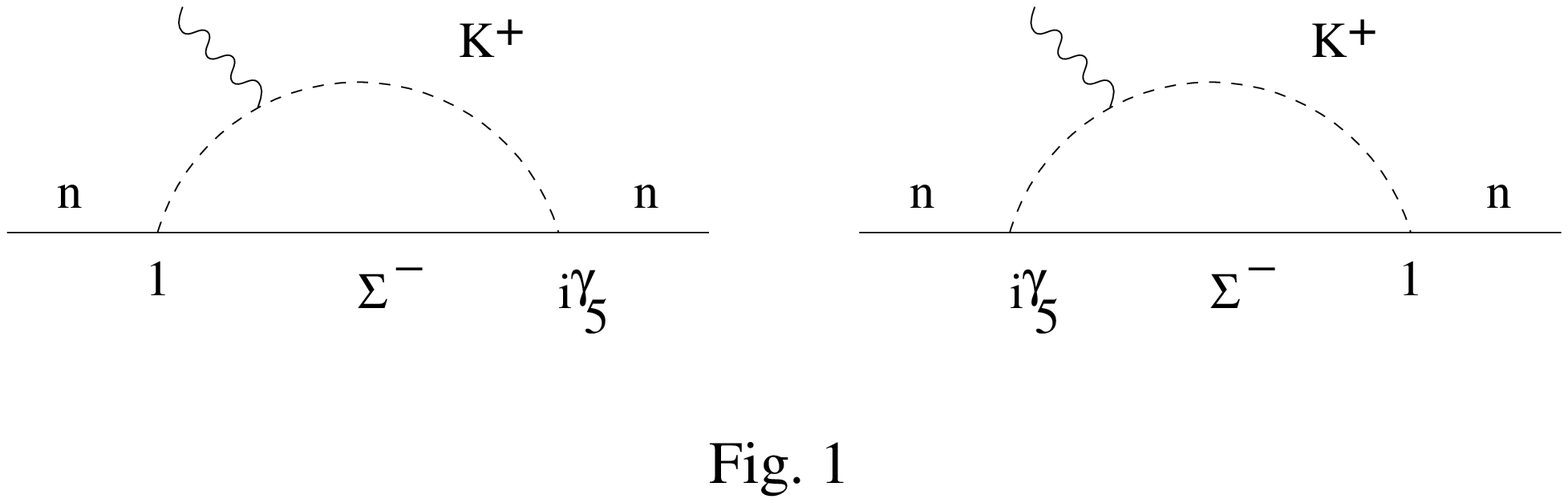}}  } 
\end{center} 
\end{figure} 
} 
 
\end{document}